\documentclass[10pt]{article}

\usepackage{microtype}
\usepackage{graphicx}
\usepackage{subfigure}
\usepackage{color}
\usepackage[hyphens]{url}
\usepackage[breaklinks=true]{hyperref}
\usepackage{fullpage}
\usepackage{booktabs}
\usepackage{authblk}
\usepackage{tabularx,array}
\usepackage{textcomp}
\usepackage[section]{placeins} % v0.87: keep floats within their section
\makeatletter
\def\fallback@sout#1{\fallback@soutwords#1 \relax}
\def\fallback@soutwords#1 #2\relax{\fallback@soutword{#1}\ifx\relax#2\relax\else\ \fallback@soutwords#2\relax\fi}
\def\fallback@soutword#1{\ooalign{#1\cr\leaders\hrule height .55ex depth -.42ex\hfill\cr}}
\IfFileExists{ulem.sty}{\usepackage[normalem]{ulem}}{\let\sout\fallback@sout}
\makeatother
\newcolumntype{L}[1]{>{\raggedright\arraybackslash}p{#1}}
\newcolumntype{R}[1]{>{\raggedleft\arraybackslash}p{#1}}

\begin{document}

%===============================================================================
% Revision notes (printed only in the marked version)
%===============================================================================

\title{Brief introduction to post-quantum cryptography in industrial applications}

%\date{v.0.93 (25/10/2026)} % this will be removed in the final version
\date{October 25, 2026}

\author[1,*]{Mateusz Żarski}
\author[1,*]{Jarosław A. Miszczak}
\affil[1]{Institute of Theoretical and Applied Informatics,	Polish Academy of Sciences, Poland}

\author[2]{Ilias Cherkaoui}
\author[3]{Cédric Crettaz}
\author[2]{Indrakshi Dey}

\author[4,5]{Adela~Georgescu}

\author[6]{Deepak V. Katkoria}
\affil[6]{Logiicdev GmbH, Graz, Austria}

\author[4,5]{Ruxandra F. Olimid}
\author[4,5]{Mihai Prunescu}
\author[7]{Maria Slim}
\author[4,5]{Silviu Laurentiu Vasile}

\author[7]{Stella Wohnig}
\affil[7]{University of Luxembourg, Luxembourg}

\affil[2]{Walton Institute for Information and Communications Systems Science, South East Technological University, Waterford, Ireland}
\affil[3]{Mandat International, Geneva, Switzerland}
\affil[4]{Department of Computer Science, University of Bucharest, Romania}
\affil[5]{Research Institute of the University of Bucharest (ICUB), Romania}

\affil[*]{Joint First Authors}

\maketitle            

\begin{abstract}
Post-Quantum Cryptography (PQC) is the main response to the threat that large,
fault-tolerant quantum computers pose to today's public-key cryptography. This
paper is a brief, self-contained introduction to PQC for practitioners in
industrial domains, with an evidence cut-off of 30 September 2026. We summarise
current resource estimates for quantum attacks on RSA, elliptic-curve and
symmetric schemes, the status of NIST standards and candidates, and the support
for PQC in major programming languages and protocol stacks. We review recent work
on PQC in four application areas that correspond to the use cases of the EU
Q-FENCE project: the Internet of Things (IoT), healthcare, civil and
infrastructure engineering, and defence. Finally, we report reproducible
micro-benchmarks of Key Encapsulation Mechanisms (KEMs) and signature schemes on
a desktop x86-64 platform with liboqs~0.15.0 and the Java~25 platform, together
with key, ciphertext, and signature sizes. We then estimate what these
objects cost at the protocol and link layers, in TLS handshakes and on constrained
radio links, summarise the requirements of the four sectors in a single table, and
discuss implementation security and how algorithm cost, data lifetime and data
sensitivity shape the choice of algorithm. Key exchange can be migrated now at
little cost, whereas authentication and constrained links remain the hard part.
The benchmarks characterise implementation cost on one reference platform; they do
not by themselves establish suitability for constrained devices.
\\[12pt]
\end{abstract}

%-------------------------------------------------------------------------------
\section{Introduction}\label{sec:intro}
%-------------------------------------------------------------------------------

Quantum computers promise to fundamentally change how we process information by exploiting the principles of quantum mechanics: superposition, entanglement, and interference~\cite{jozsa2003role}. By leveraging these effects, quantum computers are expected to deliver computing capabilities beyond the reach of any classical supercomputer \cite{arute2019quantum,ichikawa2024current}. A fully fault-tolerant quantum computer, protected by error correction, could perform long and precise computations beyond the reach of classical machines~\cite{gottesman2009introduction}.

In particular, such systems are expected to break widely used cryptographic schemes~\cite{gheorghiu2019benchmarking}. Gidney estimates that a 2048-bit RSA modulus could be factored in under a week by a machine with fewer than one million noisy physical qubits, assuming a 0.1\% gate error rate, a 1\,\textmu s surface-code cycle and a 10\,\textmu s control reaction time~\cite{gidney2025how}.

%--------------------------------------------------------------------------------
\subsection{Post-Quantum Cryptography (PQC)}
%--------------------------------------------------------------------------------

The anticipated capabilities of large-scale fault-tolerant quantum computers motivated the development of Post-Quantum Cryptography (PQC)~\cite{bernstein2009post,bernstein2017post}. PQC algorithms run on classical computers, but are specifically engineered to be secure against known attacks resulting from quantum algorithms, in addition to classical attacks. Two major examples of quantum threats include Shor's algorithm~\cite{shor1999polynomial} for public-key systems and Grover's algorithm~\cite{grover1996fast} for symmetric systems and hash functions. In particular, Shor's algorithm demonstrated that quantum computers could efficiently solve mathematical problems such as integer factorization and the computation of discrete logarithms, problems used as the basis of widely deployed public-key cryptosystems, including RSA~\cite{mosca2018cybersecurity}.

Since NIST published its first three PQC standards in August 2024~\cite{nist_fips203,nist_fips204,nist_fips205},\footnote{The standardisation of additional digital signature schemes is in its third round~\cite{nist-standard_add,nist2026round3}.} and as PQC protocols have been deployed in many popular tools and frameworks, the impact of PQC on cybersecurity has gained significant attention. During the first year after the publication of NIST standards, PQC became not only a subject of academic discussions or tests in future-facing R\&D labs~\cite{mastercard2025migration}; it is now available in major frameworks and programming languages, including OpenSSH, OpenSSL, and Java.

%--------------------------------------------------------------------------------
\subsection{Quantum Key Distribution (QKD)}
%--------------------------------------------------------------------------------
Quantum Key Distribution (QKD) is an alternative approach for safeguarding cryptography against quantum attackers. Unlike PQC, this approach uses quantum states transmitted over a quantum channel to generate key material shared by two endpoints.

QKD and PQC address different parts of the problem. QKD is a key-establishment
technique: between two nodes connected by a suitable quantum channel it can
replace a public-key key exchange such as Diffie--Hellman, but it provides neither
authentication nor digital signatures. Its classical channel must itself be
authenticated, which requires either pre-shared keys or quantum-safe signatures.
PQC, in contrast, replaces the public-key primitives used for key establishment and
for signatures with algorithms that run on conventional hardware and fit into
existing protocols.

In \cite{ravi2025quantum} an overview of QKD protocols is provided, as well as real-world implementations and challenges for deploying PQC. The paper reviews strategies for mitigating quantum threats, migration to PQC, and describes quantum-safe cybersecurity solutions, such as hybrid classical-quantum cryptographic schemes and the development of quantum-resistant protocols and standards.

QKD and PQC can be combined, for example by mixing a QKD key and a PQC-derived key into one session key; a proof-of-concept implementation of this approach was presented in~\cite{garms2024experimental}. Several national security agencies nevertheless consider QKD suitable only for niche use cases, citing the need for an authenticated classical channel, distance limits, the reliance on trusted relay nodes in practical networks, and the difficulty of evaluating the security of practical implementations, and they recommend PQC as the primary migration path~\cite{qkd_position2024}. An independent analysis of deployed QKD use cases reaches similar conclusions~\cite{aquina2025critical}, and large operators have stated the same preference~\cite{google2025qkdpqc}.

Currently, the main focus of quantum threats is asymmetric cryptography, whose security rests on the presumed hardness of integer factorisation or of the discrete logarithm problem in certain abelian groups.

%------------------------------------------------------------------------
\subsection{Scope, contribution, and novelty over related work}
%------------------------------------------------------------------------
This paper is intended as introductory material for engineers and decision makers in
industrial domains; it is not a systematic literature review. The application areas and
examples were selected to match the use cases of the EU-funded Q-FENCE
project~\cite{qfence2025}, and where we describe Q-FENCE pilots, we report their planned
functionality rather than completed evaluations. All status statements (standards,
candidates, software releases) refer to the evidence cut-off date of 30 September 2026.

We use four kinds of evidence and weigh them differently. Normative documents (NIST FIPS and
special publications, IETF drafts, and EU and national-agency guidance) define the status of
algorithms and the migration deadlines. Release notes and API documentation of software
projects establish what each ecosystem provides; we cite them with access dates because they
change quickly. Peer-reviewed papers and preprints support the attack-cost estimates and the
application review. Market reports are used only as context in Section~\ref{sec:market-analysis}.
Finally, our own measurements in Section~\ref{sec:experiments} are released with raw logs and
scripts so that they can be reproduced and extended.

The survey \cite{applicationsqpc2024} examines existing work concerning applications of PQC in various domains, including industrial applications. Unlike our work, the paper focuses on articles published in 2022-2023 and reduces to a literature review. Works that specifically refer to PQC in specific fields are further discussed in Section~\ref{sec:potential}.

We first discuss the resistance of classical schemes to quantum attacks. In the same line as our work, \cite{gheorghiu2019benchmarking} provides state-of-the-art estimates of the costs of implementing quantum attacks on the main classical cryptographic algorithms such as AES, SHA, RSA, and EC-based. However, the results in \cite{gheorghiu2019benchmarking} reflect the state of 2019 and, contrary to our work, do not benchmark PQC.

Over the years, several other papers have benchmarked PQC algorithms in different environments, including specialized hardware and resource-constrained devices \cite{hryniv2024benchmarking,performancepqc2025,turino2025pqc,gaj2018challenges,dziechciarz2024efficiency}. 
However, these differ from our work, which remains general, as discussed in Section \ref{sec:experiments}. In addition, we also look at the current support for PQC under different programming languages.

Our contributions can be summarized as follows:
\begin{itemize}
    \item We give a brief overview of published PQC market forecasts and the policy drivers behind them;
    
    \item We examine PQC essentials, including current resource estimates for quantum attacks, possible solutions, as well as PQC standardization and adoption, including the status of PQC support by programming languages;
    
    \item We discuss PQC within possible industrial applications, such as the Internet of Things (IoT), healthcare applications, infrastructure engineering, and defense;
    
    \item We summarise the requirements of the four sectors in a single table that maps data and asset lifetimes and constraints to a migration pattern;
    
    \item We provide a reproducible benchmark\footnote{Raw logs and scripts are released as accompanying material~\cite{zenodo-pqc-test-data}. } for a selected set of Key Encapsulation Mechanisms (KEMs) and signature schemes, report their key, ciphertext and signature sizes, and discuss \textit{algorithm performance}, \textit{data lifetime}, and \textit{level of data sensitivity} as selection criteria of specific PQC constructions for specific applications;
    
    \item We estimate the cost of PQC at the protocol and link layers (TLS handshakes, certificate chains and constrained radio links) and discuss implementation security and validation.
\end{itemize}
%------------------------------------------------------------------------
%\subsection{Outline}
%------------------------------------------------------------------------
The paper is organized as follows. In Section \ref{sec:market-analysis} we summarise PQC market forecasts. In Section~\ref{sec:pqc-basic} we give an overview of PQC and describe the currently used and adopted PQC solutions. In Section~\ref{sec:potential} we focus on potential application areas in several sectors, including IoT, defense, and healthcare, and describe the major existing obstacles. In Section~\ref{sec:experiments} we benchmark selected quantum-safe methods and state the limitations of these measurements. Finally, in Section~\ref{sec:final} we summarize the presented work and provide some concluding remarks.

%-------------------------------------------------------------------------------
\section{Market forecasts}\label{sec:market-analysis}
%-------------------------------------------------------------------------------

The figures in this section come from proprietary market reports. They use different market definitions (PQC products, PQC migration services, or quantum cryptography in general), so they are not directly comparable and indicate investment interest rather than technical maturity.

Currently, the market for PQC is in a state of rapid growth. Several factors drive this, including the 2024 publication of NIST standards, which enables the wide adoption of PQC solutions, and progress in the field of quantum computing, which makes \textit{harvest now, decrypt later} attacks a planning concern. Both these factors result in the market shifting from experimental research and in-house solutions towards enterprise-wide migration, also in accordance with the European Coordinated Implementation Roadmap for the Transition to Post-Quantum Cryptography~\cite{eu-pqc-roadmap}.

A recent report from Juniper Research entitled Post-Quantum Cryptography Market: 2026-2035 \cite{juniperresearch2026} estimates the growth of the global PQC market from USD 1.2 billion in 2026 to USD 13 billion in 2035, a compound annual growth rate (CAGR) of about 30\%.

The 2026--2033 edition of the Post-Quantum Cryptography Market report by Grand View Research~\cite{GrandViewResearch2024} estimates the global PQC market at USD 1.58 billion in 2025 and projects USD 20.5 billion by 2033, a CAGR of 37.8\%.

To exemplify with European initiatives, the Quantum Flagship \cite{quantumflagship2018} was launched in 2018 -- a EUR 1 billion, 10-year+ European Commission research initiative that involves research institutions, academia, industry, enterprises, and policymakers with the aim to transition quantum to commercial applications and envision the \textit{quantum web}. Similar national-level initiatives exist, for example, the Swiss quantum initiative~\cite{swissquantuminitiative}. 

Narrower market definitions give much smaller figures: MarketsandMarkets, which counts PQC products and services only, projects growth from USD 0.42 billion in 2025 to USD 2.84 billion by 2030, a CAGR of 46.2\%~\cite{marketsandmarkets2025}.

According to Future Market Insights and their report entitled Post-Quantum Cryptography (PQC) Migration Market Analysis - Size, Share, and Forecast Outlook 2025 to 2035 \cite{fmi}, the global PQC migration market is projected to grow from USD 1.9 billion in 2025 to approximately USD 12.4 billion by 2035, with a forecast to expand at a CAGR of 20.6\%. The report mentions Europe, among others, as a key growth region in the PQC migration market. The report identifies the data sovereignty requirements in Europe as an accelerating factor for the adoption of quantum-safe encryption, explicitly mentioning the EU Cyber Resilience Act and the Gaia-X PQC pilot.  

Recent observations and analysis on the quantum technology market are provided by the Global Quantum Intelligence within their Quantum Computing Report~\cite{gqi}.

In summary, the forecasts differ by almost an order of magnitude, from USD 2.8 billion by 2030 to USD 20.5 billion by 2033, with CAGRs between 20.6\% and 46.2\%, because they define the market differently. They agree only that the market will grow rapidly over the next decade.

%-------------------------------------------------------------------------------
\section{Basic concepts of PQC}\label{sec:pqc-basic}
%-------------------------------------------------------------------------------

We now introduce some PQC concepts that are important from the perspective of industrial applications. We start by describing how realistic the possibility of using quantum computers to attack current cryptographic systems is. Next, we briefly describe the proposed solution and review standardization and adoption efforts. We also describe the current state of PQC adoption in major programming languages and software development frameworks.

%-------------------------------------------------------------------------------
\subsection{Quantifying the threat}
%-------------------------------------------------------------------------------

Table \ref{tab:attacks} lists published estimates of the quantum resources required to break common cryptographic schemes. Unless noted otherwise, the estimates assume surface code error correction with $\sim$0.1\% physical gate error, a 1\,\textmu s code cycle, and planar nearest-neighbour connectivity. Here logical qubits are understood as error-corrected qubits, while physical qubits are the hardware qubits from which logical qubits are built. The rows come from different papers and cost models, so they should be compared with care.

\begin{table*}[b!]
	\centering
    \begin{small}
    \setlength{\tabcolsep}{4pt}
	\begin{tabular}{@{}llrrll@{}}
		\toprule
		\textbf{Scheme} & \textbf{Attack algorithm} & \textbf{Logical qubits} & \textbf{Physical qubits} & \textbf{Runtime (est.)} & \textbf{Ref.} \\ \midrule
		RSA-2048 & Shor (factoring) & $\approx$6{,}190 & $\approx$20\,M & 8\,h & \cite{gidney2021how} \\
		RSA-2048 & Shor (approx.\ residues) & 1{,}399 & $<$1\,M (897{,}864) & $<$1 week & \cite{gidney2025how} \\
		RSA-3072 & Shor (approx.\ residues) & 2{,}043 & not estimated & not estimated & \cite{gidney2025how} \\
		{ECC-256 (secp256k1)} & Shor (discrete log) & {$\leq$1{,}200 / $\leq$1{,}450} & {$<$0.5\,M} & {minutes} & \cite{babbush2026securing} \\
		AES-128 & Grover ({key} search) & {2{,}953} & {--} & {$\approx 2^{64}$ iterations} & {\cite{grassl2016applying}} \\
		AES-256 & Grover (key search) & {6{,}681} & {--} & {$\approx 2^{128}$ iterations} & {\cite{grassl2016applying}} \\
		SHA-256 ({preimage}) & {Grover} & $\approx$ {2{,}400} & {--} & $\approx 2^{85}$ {iterations} & \cite{gheorghiu2019benchmarking} \\ \bottomrule
	\end{tabular}        
    \caption{Published quantum resource estimates for breaking common cryptographic schemes. Estimates for Shor's algorithm assume a 0.1\% physical gate error rate, a 1\,\textmu s surface-code cycle and planar connectivity. For Grover's algorithm the dominant cost is the number of \emph{sequential} iterations, because Grover search parallelises poorly; physical-qubit counts are omitted for these rows because they depend strongly on the chosen parallelisation.}
    \label{tab:attacks}
    \end{small}
\end{table*}

Chevignard, Fouque and Schrottenloher~\cite{chevignard2025reducing} showed how to factor an $n$-bit RSA modulus with about $(0.5+\varepsilon)n$ logical qubits, at the price of many more gates; Gidney's 2025 estimate~\cite{gidney2025how} makes this trade-off practical, which explains the drop from about 6{,}200 to about 1{,}400 logical qubits in Table~\ref{tab:attacks}.

For comparison, the best classical attack on RSA-2048, the general number field sieve, is estimated at about $2^{112}$ operations, which is why NIST rates RSA-2048 at the 112-bit security level.

We remark that fault-tolerant quantum computers would break RSA and ECC with one to a few thousand logical qubits, but hundreds of thousands to millions of physical qubits are needed due to error correction. This makes them the primary vulnerable schemes. The physical-qubit count is much higher than the logical-qubit count because of the overheads of error correction, magic-state production and routing; the exact ratio depends on the hardware architecture and on the code used. Estimates also trade qubit count against circuit depth: designs with fewer qubits usually run longer. On the other hand, in the case of symmetric cryptography -- such as breaking AES and SHA -- there is only a quadratic speedup with Grover's algorithm, and this speed-up parallelises poorly: an attack on AES-128 needs about $2^{64}$ \emph{sequential} evaluations of AES inside a fault-tolerant circuit. NIST therefore uses AES-128 key search as the reference cost for its lowest security category and continues to approve AES-128~\cite{moody2024transition}, while several agencies recommend AES-256 for long-term confidentiality.

One should note that current quantum devices, described as near-term quantum computers, also called Noisy Intermediate-Scale Quantum (NISQ) devices~\cite{preskill2018quantum}, operate with hundreds to a few thousand physical qubits. Hence, they are far from the hundreds of thousands of high-quality qubits millions required to pose a serious threat for the considered protocols. However, the threat is real once scalable, fault-tolerant architectures exist.

Resource estimates for elliptic-curve attacks have fallen particularly fast. In March 2026,
Babbush et~al.\ estimated that the 256-bit elliptic-curve discrete logarithm problem can be
solved with at most 1{,}200 logical qubits and 90 million Toffoli gates (or 1{,}450 logical
qubits and 70 million Toffoli gates), which on superconducting hardware corresponds to fewer
than half a million physical qubits and a runtime of minutes~\cite{babbush2026securing}; the
authors withheld the circuits and published a zero-knowledge proof of their properties
instead; an independent, openly published construction with comparable point-addition costs
followed in June 2026~\cite{schrottenloher2026optimized}. An open benchmark, ECDSA.fail, then
invited the community to improve the dominant subroutine, a single secp256k1 point addition,
scored by the product of peak logical qubits and Toffoli count~\cite{eigenlabs,ecdsafail2026}.
By late July 2026 the best submitted point-addition circuit used 1{,}151 logical qubits and about 
1.3 million Toffoli gates, a score more than 50\% below the reference point-addition circuit,
although the two use different accounting conventions. Such rapid, partly AI-assisted progress
is a further reason to treat PQC migration as a priority.

One should note that the estimates of the resources required for cryptographically relevant attacks continue to evolve. The recently published update of the quantum-computing landscape \cite{jaques2026quantum} emphasizes that the relevant measure of progress is no longer the number of physical qubits alone but the ability to combine physical qubits, error correction, logical operations, connectivity, and computational speed into a scalable, fault-tolerant system. In particular, developments in high-rate quantum low-density parity-check (qLDPC) and related codes could substantially reduce the physical-qubit overhead traditionally associated with large-scale implementations of Shor's algorithm. However, even though experimental demonstrations of high-rate, low-overhead quantum error correction exhibit significant progress~\cite{bravyi2024high, googleqai2024threshold, wang2026demonstration}, scaling of these codes to large fault-tolerant processors remains challenging.

%-------------------------------------------------------------------------------
\subsection{Proposed solutions}
%-------------------------------------------------------------------------------

Table \ref{tab:pqc_schemes} includes a snapshot of the PQC landscape, with emphasis on the distinction between standardized schemes, schemes selected for future standardization, and research candidates.

One should note that \textit{quantum-resistant} does not mean that a scheme has been proved secure against arbitrary quantum computers. For most PQC schemes, security is based on classical hard problems for which no efficient quantum algorithm is currently known. Resource estimates therefore describe the best-known attack strategies, not guaranteed lower bounds.

\begin{table*}[b!]
  \centering
  \setlength{\tabcolsep}{2pt}
  \renewcommand{\arraystretch}{1.05}
  \begin{small}
  \begin{tabularx}{\textwidth}{L{2.1cm} L{3.9cm} L{6cm} L{4.2cm}}
    \toprule
    \textbf{Family} & \textbf{Examples} & \textbf{Best known attacks} & \textbf{Status} \\ \midrule
    Lattice-based & {ML-KEM (FIPS 203), ML-DSA (FIPS 204), FN-DSA (Falcon, FIPS 206 draft)} & {Lattice reduction (BKZ with sieving). The cost is \emph{exponential} in the block size; quantum sieving improves only the constant in the exponent.} & {ML-KEM and ML-DSA standardised (Aug 2024); FIPS 206 not yet final} \\ \midrule
    Code-based & {HQC; Classic McEliece; BIKE} & {Information-set decoding; Grover-type speed-ups reduce the exponent by at most a factor of two.} & {HQC selected for standardisation (Mar 2025)~\cite{nist_ir8545}; McEliece and BIKE not selected by NIST} \\ \midrule
    Multivariate & {UOV, MAYO, QR-UOV, SNOVA; Rainbow, GeMSS (broken)} & {Classical algebraic and MinRank attacks; Rainbow was broken classically~\cite{beullens2022}. No specific quantum attack beyond generic search.} & {Four schemes in round~3 of the additional signatures~\cite{nist2026round3}} \\ \midrule
    Hash-based & {SLH-DSA (FIPS 205); LMS, XMSS (stateful, SP 800-208~\cite{nist_sp800_208})} & {Generic preimage/collision search on the hash function; parameters already account for Grover.} & {Standardised} \\ \midrule
    Isogeny-based & {SQIsign; SIKE (broken)} & {SIKE: polynomial-time \emph{classical} key recovery (2022)~\cite{castryck2023}. SQIsign relies on different assumptions.} & {SIKE withdrawn; SQIsign in round~3} \\ \midrule
    {MPC-in-the-head} & {FAEST, MQOM, SDitH} & {Attacks on the underlying symmetric, code or multivariate problem.} & {Round~3 candidates} \\
    \bottomrule
  \end{tabularx}
    \caption{PQC families, best known attacks and standardisation status as of 30 September 2026.}
  \label{tab:pqc_schemes}
  \end{small}
\end{table*}

NIST currently has three published PQC standards, ML-KEM, ML-DSA and SLH-DSA \cite{nist_fips203,nist_fips204,nist_fips205}. FN-DSA, based on Falcon, will be published as FIPS~206 but was not final at the cut-off date~\cite{nist-standard}, and NIST selected HQC as a second KEM for standardization in March 2025~\cite{nist_ir8545}. NIST's additional-signature process is also active, with nine candidates in the third round since 14 May 2026~\cite{nist2026round3}. One of them, the lattice-based HAWK, was withdrawn by its designers on 29 July 2026 after a published key-recovery attack approximately halved the lattice-reduction block size needed to recover an equivalent secret key~\cite{nist2026round3}. The episode illustrates why migration plans should be crypto-agile even for candidates that have survived two evaluation rounds.

Supplementary to the information in Table \ref{tab:pqc_schemes}, it is worth remembering that multivariate cryptography as a broader family remains an active area of research, with newer schemes such as MAYO, QR-UOV, SNOVA, and UOV still under consideration for post-quantum digital signatures \cite{nist-standard_add}.

%-------------------------------------------------------------------------------
\subsection{Standardisation and adoption}
%-------------------------------------------------------------------------------

Recently, significant progress has been made in the adoption of the PQC standards~\cite{nist-standard}, including major operating system vendors in cloud environments~\cite{redhat-10}.
Post-quantum algorithms have recently become available in the OpenSSH 10.0 suite~\cite{openssh-10}, which made the hybrid post-quantum key exchange based on the FIPS 203 standard (mlkem768x25519-sha256) its default.

Significant progress has also been made regarding the support for post-quantum methods in major development ecosystems, confirming strong support for PQC by major industrial players. A summary of PQC support by major programming language and development environments is provided in~Table~\ref{tab:pqc-language-status-2026}.

In particular, Java Platform version 24~\cite{java-24}, released in March 2025, includes post-quantum schemes for key encapsulation and digital signatures (JEP 496 and JEP 497), making it one of the first general-purpose, widely deployed programming languages adopting PQC. 
JDK 27, released on 15 September 2026, enhances TLS 1.3 by implementing post-quantum hybrid key exchange (JEP 527), which combines quantum-resistant ML-KEM with traditional elliptic-curve algorithms to protect against future quantum attacks and provides the fundamental capabilities required for an initial transition to PQC \cite{java-27}. JDK 27 is not a long-term-support release; Oracle has announced plans to bring comparable capabilities to its LTS releases~\cite{java-27}.
In addition, recently .NET has also moved into the native-PQC category. Released in November 2025, .NET 10 provides MLKem, MLDsa, and SlhDsa, corresponding to FIPS 203, 204, and 205. On Linux, these can use OpenSSL 3.5+, while Windows support depends on the Windows cryptographic API. 
In Python, the pyca/cryptography library added ML-KEM and ML-DSA in version 47.0.0 (April 2026) for the AWS-LC and BoringSSL back ends only, extended them to OpenSSL 3.5+ in version 48.0.0, and added ML-DSA certificates in version 49.0.0 (June 2026)~\cite{pyca_changelog}. Go~1.27 (August 2026)~\cite{go127} added a public ML-DSA package integrated with its X.509 and TLS libraries, and Apple's operating systems (iOS~26 and the matching releases) provide ML-KEM and ML-DSA in CryptoKit and negotiate hybrid post-quantum key exchange in TLS by default~\cite{apple2026security}. 

\begin{table*}
\centering
\small
\setlength{\tabcolsep}{2pt}
\renewcommand{\arraystretch}{1.05}
%{p{2.4cm}|p{4.8cm}|p{4.2cm}|p{4.2cm}}
\begin{tabularx}{\textwidth}{L{2.8cm} L{5.3cm} L{4.5cm} L{3.6cm}}
\hline
\textbf{Ecosystem} &
\textbf{PQC Support} &
\textbf{Protocol integration} &
\textbf{Main limitations} \\
\hline

\textbf{Java 25 (LTS)/ JDK 27} &
ML-KEM and ML-DSA are available in the JDK through JEP 496 and JEP 497 {(since JDK 24)}. 
The KDF API is also available.
&
PQC primitives can be used through the JCA/JCE architecture. Hybrid TLS 1.3 key exchange is available from JDK 27 (JEP 527)~\cite{java-27}, not yet in the LTS release. &
SLH-DSA and some protocol and certificate integration may require external 
providers or libraries. \\
\hline

\textbf{Go 1.27} &
ML-KEM is available in the standard library (\texttt{crypto/mlkem}). Go 1.26 added HPKE and further 
PQC infrastructure; {Go 1.27 adds a public \texttt{crypto/mldsa} package~\cite{go127}.}&
Hybrid PQ TLS is enabled by default, including X25519+ML-KEM-768 and 
NIST P-curve + ML-KEM combinations. ML-DSA is supported in \texttt{crypto/x509} and in TLS~1.3 signatures~\cite{go127}. &
No SLH-DSA in the standard library; PQ certificates also depend on CA and peer support. \\
\hline

\textbf{C/C++} &
OpenSSL 3.5+ supports all three NIST standardized algorithms: ML-KEM, ML-DSA, 
and SLH-DSA. &
Native PQ and hybrid TLS groups are supported, including ML-KEM-only and 
X25519/P-curve + ML-KEM key exchanges. &
Application configuration and certificate ecosystem integration are still 
important deployment considerations. \\
\hline

\textbf{C\# / .NET 10} &
.NET 10 provides built-in support for ML-KEM, ML-DSA, and SLH-DSA through 
\texttt{System.Security.\allowbreak Cryptography} (\texttt{MLKem}, \texttt{MLDsa}, \texttt{SlhDsa}), plus composite ML-DSA signatures~\cite{dotnet10}. &
Uses platform cryptographic facilities, including Windows CNG and OpenSSL 
3.5+ on Linux. &
{\texttt{MLDsa} and \texttt{SlhDsa} are still marked experimental (\texttt{MLKem} is not), and .NET~10 offers no PQC on macOS~\cite{dotnet10}.} \\
\hline

\textbf{Rust} &
A mature ecosystem provides ML-KEM, ML-DSA, and SLH-DSA implementations, 
including independently developed and formally verified implementations. &
PQ and hybrid TLS support are available through TLS ecosystems such as 
\texttt{rustls} and specialized PQC crates. &
There is no single universally adopted PQC API equivalent to the Go or 
.NET standard-library APIs. \\
\hline

\textbf{Python} &
NIST-standardized algorithms ML-KEM and ML-DSA in \texttt{pyca/cryptography} ($\geq$47.0; OpenSSL back end from 48.0; ML-DSA certificates from 49.0)~\cite{pyca_changelog}; \texttt{pqcrypto} and \texttt{liboqs-python} for other algorithms. &
PQC TLS support is generally inherited from OpenSSL or specialized 
third-party packages. &
PQC is not part of the Python standard cryptographic API; deployment often 
depends on native libraries. \\
\hline

\textbf{JavaScript / Node.js} &
Node.js applications can access PQC through the underlying OpenSSL 
implementation and third-party libraries. &
PQ/hybrid TLS capabilities depend on the Node.js and OpenSSL versions. &
Browser WebCrypto remains substantially more limited than server-side 
Node.js cryptography. \\
\hline

\textbf{Swift / Apple platforms} &
{CryptoKit provides ML-KEM and ML-DSA (ML-DSA-65/87) from iOS/macOS~26, including Secure Enclave support~\cite{apple2026security}.} &
{Hybrid post-quantum key exchange is negotiated by default in TLS through \texttt{URLSession} and the Network framework; also in iMessage, VPN (IKEv2) and SSH~\cite{apple2026security}.} &
{No SLH-DSA; requires the 2025 OS releases, so older devices need third-party libraries.} \\
\hline

\textbf{PHP} &
PQC can be accessed through OpenSSL 3.5+ when PHP is linked 
against a suitable OpenSSL implementation. &
TLS capabilities depend primarily on the underlying OpenSSL version. &
PHP itself does not provide a comprehensive native PQC API. \\
\hline

% \textbf{Haskell / Erlang / Elixir / Julia} &
% PQC is primarily available through bindings to liboqs, OpenSSL, or other 
% external cryptographic libraries. &
% Protocol support depends on the underlying C/C++ cryptographic and TLS 
% libraries. &
% Limited native, language-level PQC APIs and greater dependence on external 
% libraries. \\
% \hline

\end{tabularx}
\caption{Support for post-quantum cryptography in major programming languages
and development environments. Status as of 30 September 2026. Where support is ``inherited'', the PQC capability of an application depends on the version of a native library (usually OpenSSL) present at run time, which a cryptographic inventory must record.}
\label{tab:pqc-language-status-2026}
\end{table*}

The European Union recognizes that the arrival of large-scale quantum computers will endanger today’s public-key cryptography. In March 2025, European Telecommunications Standards Institute (ETSI), the European standards body, announced the launch of its post-quantum security standard to guarantee the protection of critical data and communications in the future. The specification Efficient Quantum-Safe Hybrid Key Exchanges with Hidden Access Policies (ETSI TS 104 015) defines a scheme for Key Encapsulation Mechanisms (KEMs) with Access Control (KEMAC), called Covercrypt, that ensures both pre-quantum and post-quantum security~\cite{etsi2025}.

In June 2025, the Network and Information Systems Cooperation Group, established to ensure cooperation and information exchange among Member States, published the Coordinated Implementation Roadmap for the Transition to Post-Quantum Cryptography. This roadmap includes a set of recommendations to ensure a synchronized transition to PQC in all member states \cite{eu-pqc-roadmap}: Member States should start the transition by the end of 2026, complete it for high-risk use cases by the end of 2030, and migrate as many systems as feasible by 2035. 

The rapid progress in the adoption and standardization of post-quantum methods~\cite{chen2025standardisation} naturally makes them more available and, at the same time, more prone to developing new attacking techniques. In this regard, it is natural to ask if the parallel progress in the field of quantum technologies has established a serious threat to the standardized post-quantum technologies. Indeed, only by investigating possible avenues for harnessing quantum computing to assess the resilience of post-quantum protocols is it possible to ensure their security.

%-------------------------------------------------------------------------------
\section{Potential applications}\label{sec:potential}
%-------------------------------------------------------------------------------

This section focuses on selected industrial applications of PQC methods. We decided to focus on application areas that are relevant to the use cases developed within the scope of the Q-FENCE project \cite{qfence2025}. This includes recent developments concerning IoT, healthcare, infrastructure engineering, and defense. The selection is illustrative rather than exhaustive, and the Q-FENCE paragraphs describe planned pilot functionality, not completed evaluations. More information about the use cases of the Q-FENCE project can be found on the project web-site~\cite{qfence2025}.

%-------------------------------------------------------------------------------
\subsection{Internet of Things (IoT)}
%-------------------------------------------------------------------------------

The Internet of Things, as a large collection of interconnected, resource-constrained devices, offers multiple benefits, but it also opens the door to various attacks and raises many security challenges. Due to their limited computational capabilities and resource-constrained nature, IoT devices commonly rely on Elliptic Curve Cryptography (ECC), making them a direct target for quantum attacks.
Moreover, the long operational lifetime of the devices exposes them to the threat of \textit{harvest now, decrypt later}. We refer the reader to~\cite{althobaiti2020cybersecurity} for more details on the impact of quantum computers on IoT security.

Post-quantum solutions for secure key establishment suitable for the IoT landscape have recently been proposed, either as Password Authenticated Key Exchange (PAKE) with different security guarantees \cite{saeidi2025post}, tailored for mobile devices, or as Authenticated Key Agreement (AKA) for Vehicular Digital Twins \cite{cui2025post}.

Within the demanding IoT context, the most common proposal combines standardised PQC for the public-key operations with lightweight symmetric cryptography for bulk data protection.
Lightweight PQC schemes are specifically designed to secure resource-constrained devices, such as those in the IoT, against the potential threats posed by quantum computing. These schemes aim to provide robust security while minimizing computational overhead, memory usage, and power consumption.
Recently, several proposals targeting this application area have been presented. We present a classification of these schemes based on their primary objective: signature generation, KEM, and authentication. These are research proposals; unlike the NIST standards, they have not undergone public multi-year cryptanalysis.

Lightweight post-quantum \textit{digital signature schemes,} LiteQSign and MUM-HORS, were introduced in a line of recent works \cite{yavuz2025liteqsign, sedghighadikolaei2025signer}. Both signatures are multi-time, hash-based, and claimed to outperform NIST PQC standards in terms of compactness, computational overhead, and memory usage. Moreover, they both rely on the same mechanism (HORS - Hash to Obtain Random Subset) to achieve very fast signing speed, but they address the drawbacks of this mechanism through different strategies.  
Another lightweight digital signature for Digital Twins is introduced in~\cite{nouma2023post}.

Rudraksh, a lightweight post-quantum KEM based on the Learning With Errors (LWE) problem, optimized for resource-constrained devices, has been proposed in \cite{kundu2025rudraksh}. Rudraksh demonstrates significant improvements in area and frequency compared to other schemes like Kyber.

For \textit{authentication}, integration of post-quantum TLS 1.3 for Industrial IoT on Raspberry Pi hardware is shown to be feasible and is benchmarked in~\cite{Shahid2026}.

IoT is broad and encompasses several specific fields of adoption. An application targeting the automotive industry was proposed in \cite{nguyen2024lightweight}, where a lightweight cryptographic Ascon protocol~\cite{mattsson2023proposals} has been integrated with the embedded automotive environment. Ascon, now standardised by NIST for lightweight symmetric cryptography~\cite{nist_sp800_232}, is not a public-key PQC scheme; as a symmetric primitive with 128-bit keys it is affected only by Grover's algorithm.
In \cite{amiriara2025pls}, the authors proposed a post-quantum framework targeting edge computing, allowing devices to either offload intensive cryptographic tasks to a post-quantum edge server or to perform them locally.
The paper \cite{paul2022towards} proposes solutions for integrating PQC into industrial M2M communications, focusing on NIST PQC standards in existing protocols and providing symbolic security proofs. 
Finally, for the transition to PQC in the IoT, see the thesis \cite{thesissebastianpaul}, which focuses on the security analysis of applications and specific experimental setups.

\paragraph{Relevance to Q-FENCE.}
A recent example of PQC integrated into the IoT environment is the Smart Cities use-case of the EU-funded Q-FENCE project \cite{qfence2025}. The goal of this use case is to provide quantum-resistant data management for IoT devices in smart city environments, enabling secure data aggregation and real-time processing of sensitive information. The lightweight cryptographic protocol proposed by Q-FENCE aims to reduce computational loads and conserve energy without compromising security. The protocol combines PQ signatures for the secure boot mechanism with high-entropy cryptographic keys generated by a Quantum Random Number Generator (QRNG), which supplies seed entropy but does not by itself increase the security of the PQC algorithms, and lattice-based homomorphic encryption to allow computation on encrypted data streams. 

Developed within the EU-funded QUBIP project \cite{qupib}, the paper \cite{camachoruiz2026multipartner} presents the Quantum-Secure IoT-based Digital Manufacturing Pilot. This provides one of the recent examples of the integration of PQC with an IoT environment to ensure data exchange and authentication and demonstrates how post-quantum methods can be efficiently deployed across heterogeneous infrastructure with limited resources. 

%-------------------------------------------------------------------------------
\subsection{Healthcare}
%-------------------------------------------------------------------------------
Awareness of the need for secure data encryption in medicine is widespread -- \emph{e.g.}, European Health Data Space~\cite{EU_Health}, \textit{Le Ségur du numérique en santé} (The safety of digital healthcare)~\cite{FR_Health}, and the post-quantum roadmap focused on healthcare in the USA~\cite{saberikamarposhti2024post}. Health records must often be retained for decades, which places them squarely within the harvest-now-decrypt-later threat.

One of the first contributions addressing the impact of PQC on healthcare was provided in \cite{saberikamarposhti2024post}. In this paper, similarly to~\cite{RODRIGUEZMEJIAS2024104670} roadmaps, the authors, considering the challenges faced by healthcare organizations, discuss a detailed roadmap for implementing post-quantum cybersecurity solutions. The roadmap elaborates on integration issues, budget constraints, and the need for specialized training. More recently, \cite{wu2026post} provides a comprehensive pathway for healthcare organizations to transition to quantum-resistant cryptography, ensuring the long-term confidentiality of sensitive patient data while maintaining operational continuity through a staged migration strategy.

In \cite{akkal2025quantum}, the authors classify emerging quantum-based security solutions, identify key challenges, and suggest a taxonomy of Internet of Medical Things (IoMT) security techniques. Similarly, in~\cite{med_iot_usa}, the authors propose an IoT monitoring framework for intensive care units secured with the Kyber 512, standardized as the ML-KEM-512 algorithm.

In \cite{adil2025quantum}, a review of recent research on Healthcare Internet of Things (HC-IoT) applications and PQC is provided. The paper discusses both the strengths and weaknesses of the technology across various subdomains of the healthcare system. 

There is also some emphasis on medical data security in data warehouses. As data grows, much of it is stored in data warehouses not directly linked to medical units, making them vulnerable to external attacks. In~\cite{THANTILAGE2023101270}, the authors propose an effective solution for securely storing large amounts of medical data, using a range of protection methods, including post-quantum encryption, and testing various data types, data schemes, storage, security, and privacy schemes.

\paragraph{Relevance to Q-FENCE.} Within the Q-FENCE project \cite{qfence2025}, the healthcare use case aims to demonstrate increased patient privacy by securely storing and sharing medical history. The solution proposes a secure hybrid approach that combines classical and quantum resources to process encrypted healthcare data. Additionally, by using Homomorphic Encryption (HE), Q-FENCE proposes an approach to compute on patient data without decrypting it and without directly accessing sensitive data. The pilot will focus on meeting the security, privacy, and integrity requirements of the end users while adapting the solution to their needs. 

%-------------------------------------------------------------------------------
\subsection{Civil and infrastructure engineering}
%-------------------------------------------------------------------------------

One area where PQC can significantly impact public safety and health is civil and infrastructure engineering. Authors of the U.S. Homeland Security report~\cite{RR-A1367-6} emphasize that many infrastructure components have significantly extended life cycles and data storage periods, making the transition to post-quantum security systems particularly urgent.

The authors of~\cite{pqc-media} list industrial transport control systems and water and electricity supply systems as critical systems for implementing post-quantum security measures. Other studies, such as~\cite{liu2024}, also highlight the importance of implementing such methods in IoT solutions for civil engineering, given the limited resources of these systems. The need for transition is also evident in goods flow management systems~\cite{XIE20251439} and in issues related to the digitization of construction engineering, such as the implementation of Building Information Management (BIM) based systems and Digital Twins~\cite{su151310462} within the scope of on-site sensor security.

However, there are several obstacles to transitioning of security systems in this area. Among them are difficulties in adapting industrial standards for transmitting large certificates for PQC systems~\cite{RR-A1367-6} or the frequent use of outdated logic controllers in infrastructure applications, which are not suited for resource-heavy computations~\cite{liu2024}.

Considering this, the current work focuses on adapting post-quantum methods to the capabilities of available industrial solutions. The authors of~\cite{app11052082} create a hybrid system that uses a QRNG to obtain a secret key that is later used in the SPHINCS-256 protocol. In~\cite{10.1145/3638250}, the authors develop a hybrid Hardware-ASisted Efficient Signature (HASES) that provides lightweight digital signatures to be used for Digital Twins. In the field of smart grid management, authors~\cite{10738045} propose a standalone post-quantum protocol for Smart Grids, or solutions such as Quantum-Resistant Hybrid Encryption for IoT~\cite{Xiong2025} to be used on IoT devices in Smart Grid management systems.

\paragraph{Relevance to Q-FENCE.} The enterprise networks use case of the Q-FENCE project \cite{qfence2025} integrates commercial QKD within the Q-FENCE cryptographic core over a high-speed fiber network, creating a bidirectional, quantum-resistant connection to manage encryption keys across multiple sites. The solution employs both hardware-based and software-based approaches and can offer enhanced security for a variety of applications requiring high confidentiality. As discussed in Section~\ref{sec:intro}, the QKD links still require authentication, which the pilot obtains from its classical and post-quantum mechanisms. Beyond infrastructure, Q-FENCE also includes a finance-sector pilot, the Secure Multi-Party Computation for Fraud Detection pilot, which will implement a cross-bank Anti-Money Laundering (AML) system, compliant with the European directives.

%-------------------------------------------------------------------------------
\subsection{Defense}
%-------------------------------------------------------------------------------

Another field in which protection against the influence of quantum computers is particularly important is military security and defense systems.
It is already recognized \cite{nsa,white_house,dep_homeland_sec} that there is a strong need to develop and migrate to PQC systems before a quantum computer capable of breaking current encryption methods is known to be operational.
In addition, NATO has adopted a quantum strategy~\cite{nato2024}, the US Army has started a research project on a post-quantum cryptography suite for tactical networks~\cite{us_army}, and national agencies such as the French ANSSI have published transition guidance~\cite{anssi2022anssi}. The US National Security Agency's CNSA~2.0 suite requires the highest parameter sets (ML-KEM-1024 and ML-DSA-87) and stateful hash-based signatures (LMS/XMSS) for software and firmware signing in national security systems~\cite{nsa_cnsa2}.

Similarly to civil and infrastructure engineering, however, there are some challenges in implementing such solutions.
Researchers note significant performance overhead when using PQC algorithms, especially in the context of legacy systems~\cite{ahmed2025surveypostquantumcryptographysupport,hussein_2024}.
Other authors point to the need to implement a stable, multi-layered transition framework due to the concerns regarding the relatively low maturity of PQC methods and backward compatibility~\cite{geremew2025post}.

Despite the challenges, there are already several pilot implementations of PQC algorithms in the defense sector.
In~\cite{xu2024preliminary}, the authors performed an implementation study on RISC-V platforms for satellite communication networks with the consideration of usage constraints.
Similarly, in~\cite{avionics}, the authors measured the performance of PQC encryption algorithms in avionics applications.
Lastly, comparable measurements were also performed for unmanned mobile robots~\cite{varma2020postquantumsecurecommand}.

\paragraph{Relevance to Q-FENCE.} Satellite communication is essential for global connectivity, navigation, emergency services, commercial data exchange, and defence. Q-FENCE~\cite{qfence2025} therefore also develops a quantum-safe security framework tailored to satellite communication networks. The pilot secures end-to-end quantum-resistant data management across a high-latency satellite link, combining PQC, quantum-derived randomness, and quantum-assisted verification mechanisms. 

%-------------------------------------------------------------------------------
\subsection{Summary across sectors}
%-------------------------------------------------------------------------------

Table~\ref{tab:sectors} condenses the four application areas into the questions that drive a
migration plan: what must be protected and for how long, what constrains the deployment, and
which migration pattern follows. The patterns combine the standards status of
Section~\ref{sec:pqc-basic} with the measurements and protocol-level costs of
Section~\ref{sec:experiments}; they are starting points to be confirmed on the target systems,
not validated recommendations.

\begin{table*}[!htbp]
\centering
\begin{small}
\renewcommand{\arraystretch}{1.05}
\setlength{\tabcolsep}{2pt}
\begin{tabularx}{\textwidth}{L{2.0cm} L{3.2cm} L{3.2cm} L{5.8cm} L{1.8cm}}
\toprule
\textbf{Sector} & \textbf{Protected asset and timescale} & \textbf{Main constraints} & \textbf{Suggested migration pattern} & \textbf{Examples} \\
\midrule
IoT and smart cities & Telemetry with short to medium confidentiality needs; device firmware and identity for a 10--15 year service life & Radio payload and duty cycle, energy, microcontroller memory, field updates & Symmetric keys provisioned at manufacture on device links; hash-based signatures (LMS/XMSS or SLH-DSA) for firmware and secure boot; hybrid TLS from the gateway onwards & \cite{yavuz2025liteqsign,Shahid2026,camachoruiz2026multipartner} \\
\midrule
Healthcare & Patient records retained for decades, hence high harvest-now-decrypt-later exposure; integrity of medical devices & Interoperability with health information systems, retention rules, clinical safety & Hybrid key exchange (category 3 or 5) for all transport now; post-quantum key wrapping when archives are re-encrypted; ML-DSA for document signatures & \cite{saberikamarposhti2024post,adil2025quantum,THANTILAGE2023101270} \\
\midrule
Civil and infrastructure engineering & Control and sensor data; operational technology with decades-long lifetimes & Legacy logic controllers, deterministic latency, large certificates in industrial protocols, maintenance windows & PQC-capable gateways in front of devices that cannot be upgraded; hybrid VPN and TLS between sites; inventory of industrial protocols & \cite{RR-A1367-6,liu2024,10738045} \\
\midrule
Defence and satellite & Classified data, command and telemetry over long mission lifetimes & Link budget and latency, certification, long-lived trust anchors & Category-5 parameters (ML-KEM-1024, ML-DSA-87) as in CNSA~2.0; stateful hash-based signatures for firmware and commands; crypto-agile ground segment & \cite{nsa_cnsa2,xu2024preliminary,avionics} \\
\bottomrule
\end{tabularx}
\caption{Summary of sector requirements and suggested migration patterns. ``Hybrid'' combines ML-KEM with an elliptic-curve key exchange; categories refer to the NIST security categories.}
\label{tab:sectors}
\end{small}
\end{table*}

%-------------------------------------------------------------------------------
\section{Benchmarking PQC solutions}\label{sec:experiments}
%-------------------------------------------------------------------------------

For our experiments, we selected the three NIST-standardised families (ML-KEM, ML-DSA and
SLH-DSA) and, for comparison, the code-based KEMs from the fourth NIST round (Classic McEliece
and BIKE), FrodoKEM, and the multivariate signature MAYO from the additional-signature process.
Only ML-KEM, ML-DSA and SLH-DSA (together with the stateful LMS and XMSS) are listed as approved
quantum-resistant algorithms in NIST's draft transition guidance~\cite{moody2024transition}.
The experiments estimate the time required for key generation and encapsulation/decapsulation in
KEMs, and for key generation, signing and verification in signature schemes. For each scheme we
also give its NIST security category and its key, ciphertext and signature sizes.

We include tests executed using the C library \texttt{liboqs}, as well as benchmarks of the post-quantum algorithms provided by Java 25. Please be advised that \texttt{liboqs} is a research and prototyping library whose maintainers advise against using it in production~\cite{liboqs}; the results therefore characterise open-source reference and optimised implementations, not validated cryptographic modules.

A comprehensive analysis of the performance of various PQC algorithms, focusing specifically on NIST-standardized digital signature algorithms -- SPHINCS+ and Dilithium --and their practical implementations, was presented in \cite{dziechciarz2024efficiency}. However, in contrast to our work, \cite{dziechciarz2024efficiency} does not discuss KEMs and instead restricts itself to digital signatures.

The work \cite{performancepqc2025} provides results for Kyber and Dilithium and discusses the speedup of the AVX2-optimized implementation, with the aim of examining the feasibility of PQC adoption in telecom environments. Unlike our work, which targets high-level programming, \cite{turino2025pqc} evaluates PQC computational and networking performance across x86 and ARM architectures, considering a low-level benchmarking. Similarly, \cite{gaj2018challenges} looks at the implementation and benchmarking of PQC in hardware. 

%-------------------------------------------------------------------------------
\subsection{Results: C version}\label{ssec:ex_setup}
%-------------------------------------------------------------------------------

In the experiments, for each algorithm tested, we used the open-source C library \textit{liboqs}~\cite{liboqs} to prototype and experiment with quantum-resistant cryptography.
Additionally, for every test, we used the same workstation and supporting libraries to ensure test repeatability.
The parameters were as follows:

\begin{itemize}
    \item Platform -- Linux 6.8.0-85-generic (x86\_64), glibc 2.35;
    \item CPU -- AMD Ryzen 7 5800X (8 cores; AVX2 and AES-NI active);
    \item Compiler -- GCC 13.3.0, \texttt{-O3}, liboqs ``distribution'' build with run-time CPU-feature dispatch (\texttt{OQS\_\allowbreak DIST\_\allowbreak BUILD});
    \item OpenSSL -- 3.0.13 (used by liboqs for SHA-2 and AES;
    \item liboqs -- version 0.15.0, git commit \texttt{ab41146}; %(19 January 2026) with local modifications
    \item Measurement -- liboqs \texttt{speed\_kem}/\texttt{speed\_sig}; each operation repeated for 10\,s; we report the mean and population standard deviation printed by liboqs.% (runs of 20--21 January 2026)
\end{itemize}
%\note{(i) The v0.85 list described a Ryzen~7 5800X / liboqs~0.14.0 run, and the old figures came from that run, but the package contains no logs for it. This version reports the documented January 2026 run. If you prefer the 5800X run, add its raw logs and re-run \texttt{tests/make\_figures.py}. (ii) Describe the ``local modifications'' to liboqs. (iii) State the CPU frequency governor and turbo setting. (iv) Add the URL/DOI of the released code and logs.}

Table~\ref{tab:setup} lists the algorithms tested along with their NIST Security Category, the date of their first submission to NIST, and their standardisation status. The Security category indicates a security equivalent to AES-128 (1), SHA-256 (2), AES-192 (3), SHA-384 (4), and AES-256 (5). The SLH-DSA names requested by our benchmark script were resolved by the script to the round-3 SPHINCS$^+$ code, which differs from FIPS~205 in details such as domain separation. We therefore report those results as SPHINCS$^+$, and we add the FIPS~205 pre-hash variant (HashSLH-DSA-SHA2-128s with a SHA3-256 pre-hash), which liboqs~0.15.0 implements separately. 
%\note{Re-run the benchmark with the corrected harness (\texttt{tests/pqbench\_new\_build.py}, changes in \texttt{tests/pqbench\_v087\_fixes.diff}), which now requests the FIPS~205 \texttt{SLH\_DSA\_PURE\_*} variants and Falcon, reports HQC as NOT BUILT unless liboqs is built with \texttt{-DOQS\_ENABLE\_KEM\_HQC=ON}, and no longer substitutes other algorithms for unresolved names. Then replace the SPHINCS$^+$ rows in Table~\ref{tab:bench}.}

\begin{table*}[b!]
  \centering
  \begin{small}
  \begin{tabularx}{\textwidth}{>{\raggedright\arraybackslash}X c L{3.5cm} L{3.6cm}}
    \textbf{Algorithm} &
    \textbf{Security Category} &
    \textbf{First submitted} &
    \textbf{Status} \\ \hline
    \multicolumn{4}{l}{\textbf{KEM}} \\ \hline
    ML-KEM 512/768/1024                    & 1/3/5 & Nov 2017 (as Kyber) & {FIPS 203 (Aug 2024)} \\
    Classic-McEliece 348864/460896/6688128 & 1/3/5 & Nov 2017 & {Not selected by NIST} \\
    BIKE L1/L3/L5                          & 1/3/5 & Nov 2017 & {Not selected by NIST} \\
    FrodoKEM (AES/SHAKE) 640/976           & 1/3   & Nov 2017 & {Not selected by NIST} \\ \hline
    \multicolumn{4}{l}{\textbf{Signature}} \\ \hline
    ML-DSA 44/65/87                        & 2/3/5 & Nov 2017 {(as Dilithium)} & {FIPS 204 (Aug 2024)} \\
    {HashSLH-DSA-SHA2-128s (SHA3-256 pre-hash)} & 1 & {--} & {FIPS 205 (Aug 2024)} \\
    SPHINCS+-SHA2 128s/128f/192s/192f-simple (round 3) & {1/1/3/3} & {Nov 2017} & {Pre-standard version of SLH-DSA} \\
    MAYO 2/3/5                             & 1/3/5 & {Jun 2023}  & {Round-3 candidate}  \\
    \hline
  \end{tabularx}  

  \caption{List of PQC algorithms tested, provided by \texttt{liboqs} version~0.15.0. Status as of 30 September 2026}.
  \label{tab:setup}
  
  \end{small}
\end{table*}

\begin{figure}[t!]
    \centering
    \includegraphics[width=\textwidth]{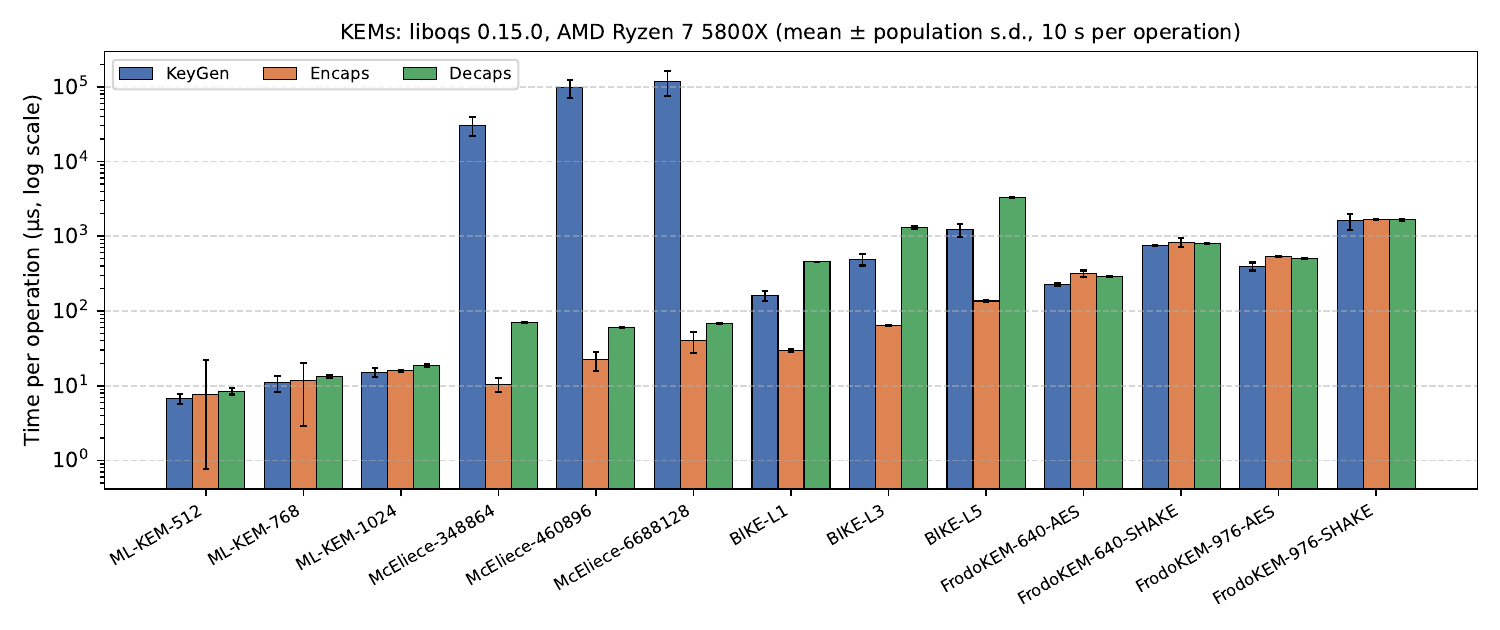}
    \caption{Benchmark results on KEM algorithms: mean time per operation with population standard deviation (error bars), liboqs~0.15.0 on an AMD Ryzen~7 5800X. Regenerated from the raw logs with \texttt{tests/make\_figures.py} script available in the accompanying materials.}
    \label{fig:kem}
\end{figure}

\begin{figure}[t!]
    \centering
    \includegraphics[width=\textwidth]{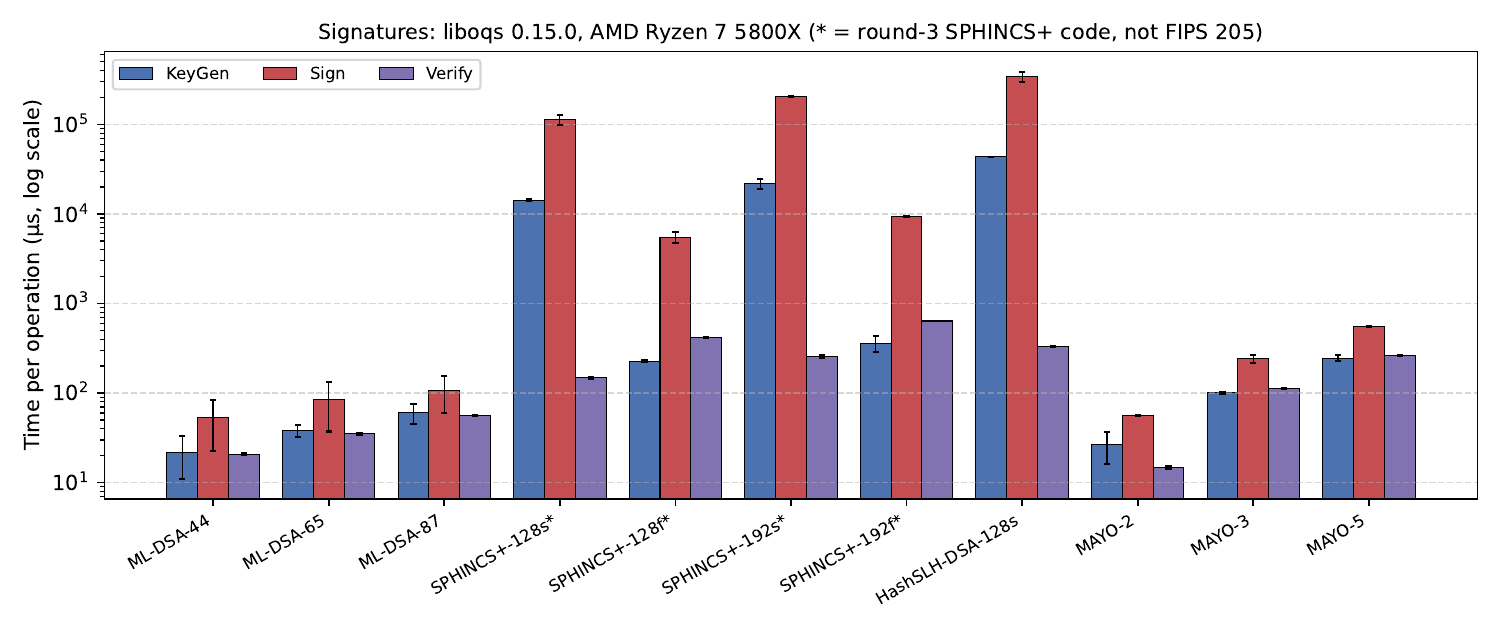}
    \caption{Benchmark results on signature algorithms, same platform and method as Fig.~\ref{fig:kem}. Entries marked $^{*}$ are the round-3 SPHINCS$^+$ implementation, not FIPS~205. The large error bars on ML-DSA signing reflect its rejection-sampling loop, whose number of iterations varies from one signature to the next.}
    \label{fig:sig}
\end{figure}

Figs.~\ref{fig:kem} and~\ref{fig:sig} list the results of our tests, and Table~\ref{tab:bench} lists the mean times together with the object sizes.
Each algorithm was tested in a 10-second time window per operation, with a different number of algorithm cycles (from 18 signing operations for SPHINCS$^+$-192s to about 666{,}000 key generations for ML-KEM-512).

\begin{table*}[t!]
  \centering
  \smallskip
  \begin{small}
  \setlength{\tabcolsep}{3.5pt}
  \begin{tabular}{@{}lcrrrrrr@{}}
    \toprule
    \textbf{Scheme} & \textbf{Cat.} & \textbf{PK (B)} & \textbf{SK (B)} & \textbf{CT/Sig (B)} & \textbf{KeyGen} & \textbf{Enc./Sign} & \textbf{Dec./Verify} \\
    \midrule
    % rows generated by tests/make_figures.py (bench_table.tex)
    ML-KEM-512 & 1 & 800 & 1,632 & 768 & 6.7~\textmu s & 7.6~\textmu s & 8.4~\textmu s \\
    ML-KEM-768 & 3 & 1,184 & 2,400 & 1,088 & 11.0~\textmu s & 11.6~\textmu s & 13.3~\textmu s \\
    ML-KEM-1024 & 5 & 1,568 & 3,168 & 1,568 & 15.1~\textmu s & 15.8~\textmu s & 18.6~\textmu s \\
    McEliece-348864 & 1 & 261,120 & 6,492 & 96 & 30.7~ms & 10.4~\textmu s & 70.5~\textmu s \\
    McEliece-460896 & 3 & 524,160 & 13,608 & 156 & 97.1~ms & 22.1~\textmu s & 60.3~\textmu s \\
    McEliece-6688128 & 5 & 1,044,992 & 13,932 & 208 & 118~ms & 39.8~\textmu s & 67.8~\textmu s \\
    BIKE-L1 & 1 & 1,541 & 5,223 & 1,573 & 161~\textmu s & 29.4~\textmu s & 454~\textmu s \\
    BIKE-L3 & 3 & 3,083 & 10,105 & 3,115 & 494~\textmu s & 64.7~\textmu s & 1.3~ms \\
    BIKE-L5 & 5 & 5,122 & 16,494 & 5,154 & 1.2~ms & 136~\textmu s & 3.3~ms \\
    FrodoKEM-640-AES & 1 & 9,616 & 19,888 & 9,720 & 227~\textmu s & 316~\textmu s & 292~\textmu s \\
    FrodoKEM-640-SHAKE & 1 & 9,616 & 19,888 & 9,720 & 753~\textmu s & 834~\textmu s & 798~\textmu s \\
    FrodoKEM-976-AES & 3 & 15,632 & 31,296 & 15,744 & 397~\textmu s & 535~\textmu s & 503~\textmu s \\
    FrodoKEM-976-SHAKE & 3 & 15,632 & 31,296 & 15,744 & 1.6~ms & 1.7~ms & 1.7~ms \\
    \midrule
    ML-DSA-44 & 2 & 1,312 & 2,560 & 2,420 & 21.9~\textmu s & 53.0~\textmu s & 20.8~\textmu s \\
    ML-DSA-65 & 3 & 1,952 & 4,032 & 3,309 & 38.1~\textmu s & 85.1~\textmu s & 35.2~\textmu s \\
    ML-DSA-87 & 5 & 2,592 & 4,896 & 4,627 & 60.5~\textmu s & 107~\textmu s & 56.4~\textmu s \\
    SPHINCS+-128s$^{*}$ & 1 & 32 & 64 & 7,856 & 14.3~ms & 113~ms & 148~\textmu s \\
    SPHINCS+-128f$^{*}$ & 1 & 32 & 64 & 17,088 & 228~\textmu s & 5.5~ms & 417~\textmu s \\
    SPHINCS+-192s$^{*}$ & 3 & 48 & 96 & 16,224 & 21.8~ms & 206~ms & 256~\textmu s \\
    SPHINCS+-192f$^{*}$ & 3 & 48 & 96 & 35,664 & 361~\textmu s & 9.4~ms & 638~\textmu s \\
    HashSLH-DSA-128s & 1 & 32 & 64 & 7,856 & 43.3~ms & 343~ms & 329~\textmu s \\
    MAYO-2 & 1 & 4,912 & 24 & 186 & 26.4~\textmu s & 56.2~\textmu s & 14.8~\textmu s \\
    MAYO-3 & 3 & 2,986 & 32 & 681 & 101~\textmu s & 242~\textmu s & 113~\textmu s \\
    MAYO-5 & 5 & 5,554 & 40 & 964 & 246~\textmu s & 548~\textmu s & 262~\textmu s \\
    
    \bottomrule
  \end{tabular}

  \caption{Object sizes (bytes, raw FIPS/specification encodings) and mean time per operation measured with liboqs~0.15.0 on an AMD Ryzen~7 5800X. For KEMs the three timing columns are key generation, encapsulation and decapsulation; for signatures, key generation, signing and verification. $^{*}$Round-3 SPHINCS$^+$ code. Standard deviations are shown in Figs.~\ref{fig:kem} and~\ref{fig:sig} and in \texttt{tests/bench\_summary.csv}.}
  \label{tab:bench}

  \end{small}
\end{table*}

As our experiments show,run times differ by up to four orders of magnitude between schemes. Within each family, cost grows with the security category, but across families speed and size are not correlated: Classic McEliece has by far the largest public keys (261\,kB to 1.04\,MB) yet one of the fastest encapsulations (19--68\,\textmu s), while its key generation is the slowest KEM operation we measured (51--228\,ms).

However, not every performance data point is equally important in certain contexts -- e.g., high keygen times for Classic-McEliece algorithms (significantly higher than the encapsulation and decapsulation time, see Fig. \ref{fig:kem}) are usually a one-time, offline cost. At the same time, encapsulation and decapsulation are still relatively fast, and the only problem in certain applications may be the large size of the public key. Finally, we note that decapsulation is generally slower than encapsulation, with a single exception for FrodoKEM (AES/SHAKE) 640/976, where the time difference is anyway low. This directly impacts the requirements on the reception side in a two-party communication system. The effect is largest for BIKE, whose decapsulation is 17--27 times slower than its encapsulation. ML-KEM is the fastest KEM in every category (15--44\,\textmu s per operation) and has moderate sizes (1.6--3.1\,kB for public key plus ciphertext). We did not evaluate parallel or hardware implementations; \cite{performancepqc2025} discusses the speed-up obtained with AVX2.

On the other hand, also in line with results obtained in \cite{dziechciarz2024efficiency,performancepqc2025}, we observe consistently asymmetric behavior in digital signatures, with the verification time being lower than the signing time. Signing cost therefore matters most where a constrained device signs often. SPHINCS$^+$/SLH-DSA signing takes between 14\,ms and 557\,ms here and produces 7.9--35.7\,kB signatures, which rules it out for frequent signing on constrained devices; it remains attractive for infrequent, long-lived signatures such as firmware, where its conservative hash-based security is an advantage. ML-DSA signs in 0.14--0.27\,ms, but its 2.4--4.6\,kB signatures, rather than computation, dominate the cost on low-bandwidth links. MAYO-2 signs as fast as ML-DSA-44 (0.14\,ms) with a 186-byte signature, at the price of a 4.9\,kB public key; it remains a candidate, not a standard.

Implementation choice matters as much as parameter choice: the FIPS~205 pre-hash implementation of SLH-DSA-SHA2-128s in liboqs~0.15.0 signs about 1.8 times slower than the optimised round-3 SPHINCS$^+$ code for the same parameter set (539\,ms versus 293\,ms), although both produce 7{,}856-byte signatures.

%-------------------------------------------------------------------------------
\subsection{Results: Java 25}\label{ssec:ex_results_java}
%-------------------------------------------------------------------------------

As an additional set of benchmarks, we include results obtained for the Java implementation of ML-KEM (FIPS 203) 
and its comparison with classical algorithms available. For the purpose of our test, we decided to use Java 25, which is, as of September 2026, Long-Term Support releases of the Java platform. Specifically, we used OpenJDK version 25.0.4 (2026-07-21) LTS. The tests were executed under Ubuntu GNU/Linux 26.04, running Linux 7.0.0-38-generic. Each test consisted of 5000 iterations, plus 200 iterations for the warmup to minimize the overhead due to interpreted bytecode execution.

As one can see from the data in Table~\ref{tab:results-java}, on the test platform, ML-KEM (FIPS 203) dominates on raw speed across all three security levels. In particular, its fastest variant ML-KEM-512 encapsulates in 88\,\textmu s and decapsulates in 56\,\textmu s, about 8 times faster than X25519 and between 80 and 120 times faster than the NIST P-curves for the same operations. Key generation is also faster for ML-KEM-512 (53\,\textmu s) than even X25519 (170\,\textmu s). However, the advantage narrows at ML-KEM-1024 (87\,\textmu s vs.\ X25519's 170\,\textmu s). These ratios describe the Java implementations, in which the elliptic-curve code is written in Java; in optimised native libraries X25519 is considerably faster.

\begin{table}[h!]
    \centering

    \small
    \begin{tabularx}{\textwidth}{L{2.7cm} L{3.1cm} R{1cm} R{1cm} R{1cm} R{1.4cm} R{1.4cm} R{1.4cm}}
        \toprule
        {Algorithm} & {Category} & {Pub (B)} & {Priv (B)} & {CT (B)} & {KeyGen (ms)} & {Encap (ms)} & {Decap (ms)} \\
        \midrule
        \multicolumn{8}{l}{\textbf{Post-Quantum Cryptography (PQC)}} \\
        \midrule
        ML-KEM-512 & PQC (FIPS 203) & 822 & 86 & 768 & 0.0526 & 0.0883 & 0.0562 \\
        ML-KEM-768 & PQC (FIPS 203) & 1206 & 86 & 1088 & 0.0642 & 0.1227 & 0.0805 \\
        ML-KEM-1024 & PQC (FIPS 203) & 1590 & 86 & 1568 & 0.0865 & 0.1859 & 0.1219 \\
        \midrule
        \multicolumn{8}{l}{\textbf{Classical Algorithms}} \\
        \midrule
        X25519 & Classical (RFC 9180) & 44 & 48 & 32 & 0.1700 & 0.6783 & 0.4752 \\
        X448 & Classical (RFC 9180) & 68 & 72 & 56 & 0.9139 & 2.6956 & 1.8636 \\
        P-256 (secp256r1) & Classical (RFC 9180) & 91 & 67 & 65 & 2.1755 & 7.3118 & 4.9800 \\
        P-384 (secp384r1) & Classical (RFC 9180) & 120 & 80 & 97 & 2.1898 & 10.8490 & 5.8136 \\
        P-521 (secp521r1) & Classical (RFC 9180) & 158 & 98 & 133 & 1.8477 & 6.9855 & 5.9102 \\
        \bottomrule
    \end{tabularx}
    \caption{Comparison of performance and key size metrics for the post-quantum and classical algorithms available in Java 25. Sizes are in bytes and times in milliseconds. Key sizes are the encodings returned by \texttt{getEncoded()} (X.509 SubjectPublicKeyInfo and PKCS\#8), not raw sizes: for example, the raw ML-KEM-512 encapsulation key has 800 bytes, and the 86-byte private key is the seed-only PKCS\#8 form. For the classical schemes, which are used as DHKEM (RFC 9180), the ``CT'' column is the encapsulated ephemeral public key.}
    \label{tab:results-java}
    
\end{table}

%\note{State the JVM vendor and version, warm-up, number of iterations and the statistic reported, and whether the Java and C runs used the same machine. (The corrected P-256/P-521 sizes are the standard DER encodings, checked with OpenSSL; Java omits the public key from EC PKCS\#8 keys, as the measured P-384 value of 80 bytes confirms.)}

The primary tradeoff is bandwidth. For example, ML-KEM-768's 1206-byte public key and 1,088-byte ciphertext are roughly 27× and 34× larger than X25519's 44-byte key and 32-byte ciphertext. This makes ML-KEM well-suited for latency-sensitive, low-round-trip contexts (TLS handshakes, local KEM) but less ideal for bandwidth-constrained channels. Within the ML-KEM family, each step up in security level (512→768→1024) costs about 20--50\% more time per operation and 30--50\% more key size, a predictable and manageable scaling.

%-----------------------------------------------------------------------------
\subsection{From primitives to protocols: bytes on the wire}\label{ssec:wire}
%-----------------------------------------------------------------------------

Micro-benchmarks measure computation, but deployed protocols also pay in bandwidth, and on many
industrial links bandwidth is the scarcer resource. In TLS~1.3 with the hybrid group
X25519MLKEM768, the client's key share grows from 32 to 1{,}216 bytes and the server's from 32 to
1{,}120 bytes~\cite{ietf_ecdhe_mlkem}. A ClientHello carrying such a share no longer fits in a
single small packet, which middleboxes that expect a one-packet ClientHello may mishandle.
Authentication grows more than key exchange: an ML-DSA-44 certificate carries a 1{,}312-byte
public key and a 2{,}420-byte signature, so a chain of two certificates plus the handshake
signature adds about 9.9\,kB, compared with roughly 350 bytes for ECDSA P-256.

Table~\ref{tab:links} converts these object sizes into link-layer frames for three technologies
common in industrial IoT. One ML-KEM-768 encapsulation key needs 15 IEEE~802.15.4 frames, 60
Bluetooth Low Energy notifications with the default MTU, and 24 LoRaWAN frames at the slowest
European data rate, where the 1\% duty cycle stretches its transfer to about 1.8~hours. Compact
signatures help: a Falcon-512 signature needs a third of the frames of an ML-DSA-44 signature,
and the MAYO-2 candidate a tenth. No standardised KEM, however, has keys or ciphertexts below
about 700 bytes. The figures ignore fragmentation headers, retransmissions and acknowledgements,
so real costs are higher. For low-power wide-area networks, public-key PQC is therefore better
placed at the gateway or network server, while devices rely on symmetric keys provisioned at
manufacture, which Shor's algorithm does not threaten.

\begin{table*}[!htbp]
\centering

\smallskip
\begin{footnotesize}

\setlength{\tabcolsep}{4pt}
\begin{tabular}{@{}lrrrrrr@{}}
\toprule
\textbf{Object} & \textbf{Bytes} & \textbf{802.15.4} & \textbf{BLE} & \textbf{LoRa DR5} & \textbf{LoRa DR0} & \textbf{Time at SF12, 1\% duty} \\
\midrule
X25519 public value & 32 & 1 & 2 & 1 & 1 & 2\,s \\
ECDSA P-256 signature (raw) & 64 & 1 & 4 & 1 & 2 & 5\,min \\
ML-KEM-768 encapsulation key & 1,184 & 15 & 60 & 6 & 24 & 1.8\,h \\
ML-KEM-768 ciphertext & 1,088 & 14 & 55 & 5 & 22 & 1.6\,h \\
X25519MLKEM768 client share & 1,216 & 16 & 61 & 6 & 24 & 1.8\,h \\
ML-DSA-44 signature & 2,420 & 30 & 121 & 11 & 48 & 3.6\,h \\
ML-DSA-44 public key & 1,312 & 17 & 66 & 6 & 26 & 1.9\,h \\
FN-DSA-512 (Falcon) signature, max. & 752 & 10 & 38 & 4 & 15 & 1.1\,h \\
SLH-DSA-128s signature & 7,856 & 97 & 393 & 36 & 155 & 12.0\,h \\
MAYO-2 signature (candidate) & 186 & 3 & 10 & 1 & 4 & 14\,min \\
\bottomrule
\end{tabular}
\end{footnotesize}

\caption{Link-layer frames needed to carry one object (analytical, computed with \texttt{tests/link\_frames.py}). Payload per frame: IEEE~802.15.4, 81 bytes (127-byte frame with maximal MAC header and AES-CCM-128); BLE, 20 bytes (default ATT MTU); LoRaWAN EU868, 222 bytes at DR5 (SF7) and 51 bytes at DR0 (SF12). The last column is the time to send the object at SF12, 125\,kHz and coding rate 4/5 under the 1\% duty cycle. Upper-layer and fragmentation headers, retransmissions and acknowledgements are ignored.}
\label{tab:links}
\end{table*}

%-----------------------------------------------------------------------------
\subsection{How to choose post-quantum solutions?}
%-----------------------------------------------------------------------------

Considering the above results, we can define the main criteria for selecting the appropriate algorithm for a specific application:

\begin{itemize}
	\item \textbf{Algorithm performance} can be crucial for lightweight web, IoT, or Remote Procedure Calls (RPC) solutions where sessions are short and data is not stored, or when communicating edge devices have low performance. On constrained links, the size of keys, ciphertexts and signatures usually matters more than computation time.

	\item \textbf{Data lifetime} is an important factor when the data being transferred has a long shelf life, e.g., medical data, personal data, or organizational data, especially when it can be read and further decrypted under the harvest now, decrypt later paradigm. Following Mosca~\cite{mosca2018cybersecurity}, data are at risk when the time they must stay confidential plus the time needed to migrate exceeds the time until a cryptographically relevant quantum computer exists; such data need quantum-safe key establishment now.

	\item \textbf{Level of data sensitivity} significantly impacts the choice of an appropriate algorithm. The higher the sensitivity of the data, the higher the enforced security. For classified or long-lived sensitive data this means the highest security categories, as in CNSA~2.0 (ML-KEM-1024, ML-DSA-87)~\cite{nsa_cnsa2}, together with crypto-agile designs that allow algorithms to be replaced.
\end{itemize}

In practice these criteria lead to a small number of default profiles. For new deployments of
transport security, hybrid key establishment that combines ML-KEM with an elliptic-curve
exchange (for example, X25519MLKEM768 in TLS~1.3~\cite{ietf_ecdhe_mlkem}) protects against
harvest-now-decrypt-later while hedging against weaknesses in the new algorithms. ML-DSA is the
default general-purpose signature; SLH-DSA or the stateful LMS/XMSS schemes~\cite{nist_sp800_208}
suit long-lived firmware and code signing; and FN-DSA is a candidate for size-critical signatures
once FIPS~206 is final. For constrained devices, the choice should be confirmed by measurements on
the target hardware.

Other selection criteria include the flexibility of the algorithms, e.g., algorithms that accept various keys/ciphertexts/signatures/etc. sizes or allow good hardware speedups (also strongly related to algorithm performance). Another important factor in real-life implementations is the side-channel resistance, which can bypass the pure mathematical security of a given algorithm and is related to the level of data sensitivity that must be maintained regardless of the implementation \cite{performancepqc2025,alagic2024status,nist_ir8545}. 

%-----------------------------------------------------------------------------
\subsection{Implementation security and validation}\label{ssec:implsec}
%-----------------------------------------------------------------------------

The mathematical security of a scheme is necessary but not sufficient; in practice, the
following implementation aspects often decide whether a deployment is secure.
\begin{itemize}
\item \textbf{Constant-time code.} Secret-dependent timing, for example from divisions in early
implementations of the Kyber reference code, has been shown to leak secret keys. Implementations
should be constant-time, and formally verified code is now available for ML-KEM in several
ecosystems.
\item \textbf{Decapsulation failures.} ML-KEM uses the Fujisaki--Okamoto transform with implicit
rejection; an implementation must not reveal, through errors, timing or logs, whether a
ciphertext was rejected.
\item \textbf{Randomness.} Key generation and hedged ML-DSA signing need a sound random number
generator. A quantum random number generator, as in the Q-FENCE pilots, should feed a standard
deterministic random bit generator rather than replace it.
\item \textbf{Physical attacks.} Embedded devices need countermeasures such as masking against
power and electromagnetic side channels, and fault-injection protection for the rejection loop of
ML-DSA and for hash-based signatures.
\item \textbf{Stateful signatures.} LMS and XMSS are secure only if no one-time key is ever reused,
so their state must be managed, typically inside a hardware security module; this makes them
suitable for firmware signing but not for general use.
\item \textbf{Validation.} Regulated products need implementations validated against official
test vectors and, where required, FIPS~140-3 module validation. Our own harness illustrates the
pitfall: it counted a known-answer-test program that ran, or even one that reported the algorithm
as not compiled, as a pass, and it substituted pre-standard code for unresolved algorithm names.
Both errors have been fixed in the released scripts.
\item \textbf{Crypto-agility.} Algorithm identifiers should be negotiable, trust anchors
updatable, and a machine-readable cryptographic inventory, including the versions of native
libraries from which many ecosystems inherit their PQC support (Table~\ref{tab:pqc-language-status-2026}),
should be kept current.
\end{itemize}

%-----------------------------------------------------------------------------
\subsection{Limitations}
%-----------------------------------------------------------------------------

The benchmarks have several limitations. They were run on desktop-grade systems with x86-64 processors,
so absolute times will differ on other hardware, and they say nothing about
memory footprint, code size or energy, which often decide feasibility on embedded devices. They
measure isolated operations, not complete protocols; the protocol- and link-level costs in
Section~\ref{ssec:wire} are analytical estimates, and handshake latency was not measured. Additionally, \texttt{liboqs} is not a validated
module, and the hash-based results use the round-3 SPHINCS$^+$ code rather than FIPS~205
except for the single pre-hash variant reported. Finally, the Java measurements come from a
different software stack, developed with different usage scenarios in mind, and are not directly comparable with the C results.

%-------------------------------------------------------------------------------
\section{Conclusions}\label{sec:final}
%-------------------------------------------------------------------------------

Published resource estimates for quantum attacks on RSA and elliptic-curve cryptography fell by more than an order of magnitude between 2019 and 2026, and policy roadmaps now set migration deadlines independently of any forecast of when a cryptographically relevant quantum computer will exist: the EU roadmap asks for high-risk use cases to be migrated by the end of 2030~\cite{eu-pqc-roadmap}, and NIST's draft transition guidance deprecates 112-bit RSA and elliptic-curve schemes after 2030 and disallows them after 2035~\cite{moody2024transition}. This makes quantum-resistant encryption methods essential to counter future risks and prevent potential harvest now, decrypt later (HNDL) attacks.

The real value of post-quantum protocols depends on secure, efficient implementation tailored to specific devices and use cases. Good PQC should be robust, scalable, purpose-fit, and designed from the ground up to balance security with performance and efficiency requirements. With RSA and ECC set to be phased out by 2035 in the NIST guidance, organizations must act now, embedding PQC into products and supply chains to ensure compliance, interoperability, and crypto agility.

In this regard, edge cases like ultra-small embedded devices and high-speed networking modules highlight key challenges. They require implementations designed with the requirements of size, performance, and long-lifecycle security in mind. 

Our current study investigated post-quantum implementations from various perspectives, including market forecasts, standardization and adoption, and possible industrial applicability. We also discussed the status of PQC implementation support by programming language. Finally, we provided a reproducible benchmark for a selected set of KEMs and signature schemes, which aligned with previous (partial) findings, and identified the main criteria for selecting an appropriate algorithm for a specific application.

Three practical conclusions follow from this study. First, key exchange can be migrated now: hybrid ML-KEM key exchange is available, often by default, in the main TLS stacks and costs tens of microseconds and about two kilobytes per handshake. Second, authentication is the harder part of the migration, because larger signatures and certificates affect public-key infrastructures, firmware signing and every relying party. Third, on constrained links the size of keys, ciphertexts and signatures, rather than computation time, is the binding constraint, which favours gateways that terminate post-quantum protocols on behalf of constrained devices.

%-------------------------------------------------------------------------------
\section*{Acknowledgment}
%-------------------------------------------------------------------------------
This work was partially supported by EU grant Q-FENCE ``Securing Tomorrow’s Digital Infrastructure with Quantum-Resistant Cryptography'' (project number 101225708) and innovation programme under grant agreement No. 101225708 as well as the Swiss State Secretariat for Education, Research and Innovation (SERI).

Authors would like to thank Eros Camacho-Ruiz and Antonio Lioy for helpful remarks and interesting discussions.
% \note{Both recommended venues (MDPI and IEEE) require authors to disclose the use of AI tools in preparing the manuscript; add the statement here.}

% ---- Bibliography ----
\bibliographystyle{abbrvurl}
\bibliography{pqc_applications_review}

\end{document}